\documentclass[a4paper,twoside]{article}
\newcommand{\affil}[1]{$^{\rm #1}$}

\usepackage[authoryear]{natbib}
\bibpunct{(}{)}{;}{a}{}{,}
\usepackage{graphicx}
\date{} 
\title{\large\bf\flushleft Future Directions in Astronomy Visualisation }
\author{\parbox{\textwidth}{\flushleft
\vspace{-0.5cm}
{\it C. J. Fluke\affil{A,B}, P. D. Bourke\affil{A}, and D. O'Donovan\affil{A}}\\
\vspace{0.4cm}
{\small \affil{A}\,Centre for Astrophysics \& Supercomputing, Swinburne University of Technology, PO Box 218, Hawthorn, Victoria 3122, Australia}\\
{\small \affil{B}\,Email: cfluke@swin.edu.au}}}
\begin{document}
\twocolumn[
\begin{changemargin}{.8cm}{.5cm}
\begin{minipage}{.9\textwidth}
\vspace{-1cm}
\maketitle
%
%
\small{\bf Abstract:}

Despite the large budgets spent annually on astronomical research equipment such as 
telescopes, instruments and supercomputers, the general trend is to analyse and view 
the resulting datasets using small, two-dimensional displays.  We report here 
on alternative advanced image displays, with an emphasis on displays that we have 
constructed, including stereoscopic projection, multiple projector tiled displays 
and a digital dome.  These displays can provide astronomers with new ways of 
exploring the terabyte and petabyte datasets that are now regularly being produced 
from all-sky surveys, high-resolution computer simulations, and Virtual Observatory
projects.  We also present a summary of the Advanced Image Displays for Astronomy (AIDA) 
survey which we conducted from March-May 2005, in order to raise some issues 
pertitent to the current and future level of use of advanced image displays.


\medskip
\medskip
\end{minipage}
\end{changemargin}
]
\small

\section{Introduction}
Astronomy is possibly the most visual of all the sciences, in both the way 
the data is collected and analysed.  Optical telescopes take images of 
the night sky so that the position, orientation, size, shape, brightness 
and colour of celestial objects can be determined (Fomalont 1982).  
Radio telescopes record intensity, polarisation and velocity data that 
is converted into pseudo-colour images or 3D spectral line cubes.  Numerical simulations 
produce datasets that are often inspected visually before being compared 
statistically with surveys. Data reduction, a key step in the analysis 
of astronomy data, is best performed by eye---the human brain has 
incredible pattern matching abilities that are yet to be reproduced 
with a computer algorithm (e.g. Norris 1994; Gooch 1995).  Visual 
representations allow the user to see patterns and relationships that are not 
apparent in simple lists of numerical results (Domik \& Mickus-Miceli 1992).  

Each year, astronomers spend millions of dollars on research equipment: 
telescopes, satellites, instruments and supercomputers. Yet the general 
trend is to 
analyse and explore the resulting observational and numerical datasets on 
small (e.g. 17'' to 21'' diagonal) two-dimensional computer monitors. 
Current and future facilities (e.g. Square Kilometre Array, Large Scale 
Synoptic Telescope), surveys (e.g. 2 Micron All Sky Survey, Sloan Digital 
Sky Survey, Gaia) and supercomputer simulations (e.g. where $N>10^8$ particles) 
provide datasets measured in terabytes and petabytes.  Virtual Observatory 
projects are bringing together disparate data archives for the 
research community to explore---with millions of objects, each having
multiple parameters, it is an increasingly complex task to make 
sense of these volumes of data (Welling \& Derthick 2001).  Standard
visualisation techniques using small, two-dimensional displays cannot 
hope to provide astronomers with a complete understanding of relationships,
dependencies, and spatial features over a range of resolutions and length/size
scales in complex $n$-dimensional datasets.   Data mining 
techniques, whether machine-oriented or human-directed, are becoming ever 
more important (e.g. Teuben et al. 2001; Mann et al. 2002; Beeson, 
Barnes, \& Bourke 2003). 

With cost savings and graphics performance driven by the consumer markets
for computer games\footnote{e.g. improved polygon drawing rates and on-board
memory on graphics cards} and home theatre,\footnote{e.g. significant reductions
in cost, increasing brightness, resolution and dynamic range, and broad product availability} it has become 
feasible to produce affordable advanced image displays such as high-resolution tiled displays, stereoscopic 3D projection 
and digital domes, with commodity or `off-the-shelf' components.  
We have witnessed on-going increases in computing 
and graphics power, and developments in `state of the art' image displays 
which have affected both single-user systems (e.g. higher resolution 
monitors, flat-panel plasma and autostereoscopic displays) and 
collaborative visualisation environments.

Since the review of display techniques and image analysis by Rots (1986), 
there has been no systematic investigation into the usefulness of advanced 
display technologies for astronomical datasets [although for specific cases 
see Norris (1994), Hultquist et al.\ (2003), Joye \& Mandel (2004)].  
We have reached a stage where advanced image displays can start to be 
more useful to, and more widely used by astronomers. Therefore, it is 
an appropriate time to take stock of the tools that are now available, 
assess their value to astronomers, and look ahead to future techniques 
for visualising datasets of increasing complexity.
Our emphasis is on non-standard image display devices.  This allows us 
to explore possibilities beyond the conventional 2D techniques 
(paper, computer monitors, overhead projectors and data projectors).

This paper is set out as follows.  In sections 2 and 3, we provide
descriptions of a number of advanced image displays, focusing 
on the specific systems that we have constructed as demonstrators:
stereoscopic projection, the digital dome, multiple projector tiled
displays and the Virtual Room.
In section 4, we report on some of the issues raised by an informal survey 
on the level of awareness of advanced images displays that was targeted 
at members of the Astronomical Society of Australia.  Finally, in section 5, 
we look at some of the limitations that advanced image displays and 
visualisation systems most overcome before they can be more widespread 
among the astronomical community.  

\section{Two-Dimensional Displays}
In this section we investigate two types of 2D display devices: large format
tiled displays and digital domes.  Both solutions can provide very high 
resolution display environments, that are ideal for collaborative
investigations or group presentations.  

\subsection{Large Format Tiled Displays}
\label{tile}
\subsubsection{Overview}
CCD detectors are now well over the 10k $\times$ 10k = $10^8$ pixel 
limit (e.g. CFHT-Megaprime, SDSS, MMT-Megacam, and the upcoming VISTA
telescope).  Typical CRT/LCD monitors for desktop and notebook computers 
have resolutions ranging from 1024 x 768 (XGA) to 1600 x 1200 (UXGA), with 
XGA also proving popular (and affordable) for data 
projectors.\footnote{The favoured XGA resolution is partly controlled by the
growing home theatre market, as this is most compatible with the PAL,
NTSC, and High Definition television standards.  At the time of writing, 
advances in digital cinema are leading to projectors with resolutions of
4096x2160 pixels, but at very high cost.}
Therefore, there are many more pixels in the datasets than a single monitor
or projector can handle---in order to see the full picture, low levels of 
detail must be omitted.  To see the dataset at full resolution, only part
of the image can be viewed at one time and the user must pan, roam or 
rely on their memory to see the large-scale features (Welling \& Derthick 
2001).

While it might seem that increasing the pixel resolution of monitors is
a solution, there is a physical limitation:  the angular resolution of the
eye is about 0.02 deg $\sim 1/3000$ radians (Fomalont 1982 and references 
therein).  For a viewing distance, $d$ m, pixels smaller than about 
$p \sim d/3000$  m will not be resolvable.  Now $p = x/N_x$ where 
$x$ is the horizontal screen diameter, and $N_x$ is the horizontal 
pixel resolution.  For example, for a laptop with $x = 0.25$ m and 
$d = 0.5$ m, there is no real benefit in going beyond a horizontal pixel 
resolution of $\sim 1500$ pixels.  To get benefit from more pixels, the 
viewer would need to move closer to the screen, which is neither practical
nor comfortable for a small screen.  Alternatively, the display can be made 
much larger, by using one or more data projectors.

\subsubsection{Our Solution}
While it is possible to purchase data projectors with a tiling capability
built-in, our approach was to use lower-cost, commodity projectors, 
driven by different computers or different graphics pipes on the same
computer.  Since commodity projectors are not designed with tiling in mind, 
it is not always possible to reliably align multiple projectors in a
pixel perfect way.  A gap between the images or a double bright seam is the
usual visual result, both of which are not ideal for content where the
virtual camera is panning or objects are moving across the seam.  These
problems can be overcome by overlapping the two images, and modifying the
pixels in the overlap region to reduce the overlap's visibility.  Projector
misalignments or lens aberrations will now only be seen as a slight blurring
of the image, and not as a sharp seam or gap.

A simple blending approach is to fade the intensity of the images to black
within the overlap region.  This approach works equally well for any number of 
images and also for images that may not be aligned in a rectangular fashion. 
Using a pair of XGA DLP projectors, we produced a 1792 x 768 pixel 
tiled display with a 256 pixel-wide overlap.  
The degree of overlap is dictated by the amount of gamma correction required 
(see below) and the dynamic range of the blend function.\footnote{In our 
initial testing, we found that the preferred 128 pixel overlap was not 
sufficient.  Powers of 2 in overlap size are not critical, but 
they simplify the programming model using OpenGL textures.}

We use a blending function, f($x$), of the form:
\begin{eqnarray}
f(x) & = &  \frac{1}{2}(2 x)^p  \,  \mbox{for $0 \le x \le 0.5$} \nonumber \\
     & = & 1 - \frac{1}{2}\left[2 (1 - x)\right]^p \,  \mbox {for $0.5 \le x \le 1$}  \nonumber
\end{eqnarray}
although a range of blend functions are possible.  For simplicity, we 
normalise the pixel coordinates of the blend region to be $0 \leq x \leq 1$.  
The exact curvature of the blend function is controlled by the 
parameter $p$. Blending is linear for $p=1$, although this tends to
result in a visible step at the edge of the blending region.  The 
transition around $x = 0.5$ becomes steeper as p increases, and we have
found $p=2$ to be a reasonable choice.  For each pixel in the overlap 
region, the final pixel value is the sum of the right image pixel 
value multiplied by f($x$) and the left image pixel value multiplied by 1-f($x$).

The blending function is implemented as a gradient mask applied to an OpenGL
texture, however, the mask on its own does not produce the correct 
blending.  Instead, a grey band appears within the overlap region. 
This is because we are adding pixel values when we should be adding
brightnesses, which can be achieved by compensating for the display gamma. 
The output brightness (normalised in the range 0 to 1) is
the pixel value raised to the power of $G$, usually in the range $1.8 \le
G \le 2.2$.

Fortunately, this is readily corrected by applying an inverse gamma power. 
The total transformation of the image pixels is f($x$)$^{1/G}$ and 
f($1-x$)$^{1/G}$ for the two image streams. In general, the gamma correction 
needs to be applied to each r, g, b value separately.  

A limiting factor in any approach is the degree to which the 
projector can create black.  While CRT projectors produce the best 
black, they are undesirable for other reasons (bulk, calibration, low 
light levels).  LCD projectors typically have very poor black levels, 
and DLP projectors are somewhere in between.

\subsection{Digital domes}
\subsubsection{Overview}
To the ancient astronomers, the night sky was an enormous sphere 
rotating around the Earth.  Although our world-view has changed dramatically, 
this spherical model is still very convenient to use.  It is somewhat 
surprising that astronomers display maps of the night sky (e.g. 
all-sky surveys of pulsars, galaxy maps or clouds of neutral hydrogen) 
on small, flat, low angular-coverage monitors using mapping techniques that 
distort areas and spatial relationships.\footnote{Consider the Mercator 
projection common for maps of the Earth.  This mapping of the spherical 
Earth to a 2D  surface does not preserve area, so that Polar countries 
like Greenland appear highly distorted. }  

The exception is the astronomy education world, where planetarium domes 
provide an idealised representation of the sky.  Planetarium projection
technology has come a long way since the world's first opto-electric 
projector was constructed by Zeiss Optical Works for the Deutsche Museum 
in Munich.  In part, these advances have been driven by consumer 
desires with viewers exposed to more sophisticated animations 
on television than planetariums could present 
(Murtagh 1989).  During the last decade, a mini-revolution 
has occurred with the emergence of full-dome video systems, available 
from a growing number of vendors.\footnote{For example, Evans \& Sutherland,
Konica Minolta, Silicon Graphics, Sky-Skan, Inc. and Spitz Inc.}
Typically 5--7 projectors display computer-generated, edge-blended content 
that is projected onto the entire dome at resolutions up to 4000 x 4000 
pixels.\footnote{Note that this is the size of square dome frames---the
actual projected area is a maximally inscribed circle, so that $\sim 21.5\%$ 
of pixels are unused. }

With the notable exceptions of the Hayden Planetarium at the American Museum 
of Natural History, New York, which has been used to visualise 
astronomical surveys in the Digital Universe project (Abbott et al. 2004) 
and large-scale numerical simulations (Teuben et al. 2001), and the
Cosmic Atlas project of the Gates Planetarium in  Denver, planetarium 
domes have been under-utilised as data exploration environments.  Reasons 
for this include:

\begin{itemize}
\item Availability and accessibility. Fixed installations require a great 
deal of physical space, leading to their placement in museums and 
science centres 
away from researchers;
\item Limited dataset size.  Traditional opto-electrical star projectors 
could not show generic datasets, and the first generation of 
digital star projectors that appeared in 1990s were limited to datasets of a 
few 1000 particles; 
\item Low resolution/low definition. Early digital solutions 
suffered from noticeable image distortions (e.g. non-uniform pixel sizes, 
so that digital stars near the horizon are stretched), and projected 
in monochrome; 
\item Lack of software tools. Designed to integrate with other 
planetarium show playback components, these systems do not use formats 
that astronomers are more experienced with; and
\item Cost. A full-dome projection system plus large ($\geq 10$ m) dome 
can cost well over \$1 million.  Unless the system was to be in nearly 
constant use for scientific visualisation, the expenditure is extremely hard 
to justify.
\end{itemize}

The next step in digital dome projection is just occurring:  a range of 
single projector solutions are entering the market, many of which use the 
angular fish-eye lenses designed by the 
elumenati.\footnote{http://www.elumenati.com}  Coupled with this is the
growing availability of portable, inflatable domes that are light, easy 
to set-up and pack away. 

Unlike a normal fish-eye lens, the elumenati lens produces a constant pixel 
size across the dome.  These lenses are still quite expensive, and are 
only suitable for a limited range of projectors.  This means that if an 
upgraded projector becomes available (e.g. with increased pixel numbers, 
or a larger dynamic range from black to white), a new lens must also be 
purchased - assuming that a fish-eye designed for that projector exists.  
Another limitation is that the lens is usually placed in the
centre of the dome, however, for a small dome, this is often the ideal
viewing position.  

\begin{figure}

\includegraphics[width=7.5cm]{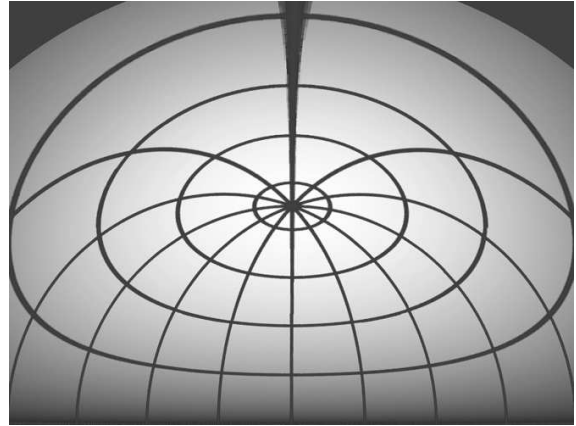}
\caption{A warped polar grid pattern ready for projection onto a dome 
surface using a spherical mirror.}
\label{domewarp}

\end{figure}

\subsubsection{Our Solution}
We have developed an alternative single projector solution that 
does not require a fish-eye lens.  Our approach uses a spherical mirror 
to reflect images from the projector onto the dome surface.  The mirror 
is placed at the edge of the dome, freeing up the centre for the viewer.

A polar grid projected with a spherical mirror onto a dome 
with a standard DLP projector (image aspect ratio of 4:3) will 
appear distorted:  equally-spaced lines of latitude will not be parallel
with the dome horizon line, and will tend to `bunch-up' close to the mirror
location.  To create an undistorted image the projected image needs to 
be pre-distorted, a process we refer to as `warping', as shown in 
Figure \ref{domewarp}. 
Displayed on the dome,  the image now has the pole at the dome's zenith, 
and latitude lines are parallel with the horizon.  

There are three ways of deriving the warping transformations: 
\begin{itemize}
\item The mapping can be derived analytically. While this may be possible 
for idealised arrangements, for other more real world situations it can be 
cumbersome; 
\item Develop an application that allows the mappings to be created 
interactively, by moving vertices and (u,v) coordinates of a mapping mesh 
until the correct mapping is achieved; or 
\item Simulate the projection environment by tracing rays from the projector 
through each pixel in the image, reflect the ray off a virtual mirror 
and onto the dome. Once the position on the dome is known the mapping for 
the pixel in question can be calculated.  We have found this method to be
the most useful for an arbitrary dome, mirror and projector configuration.
\end{itemize}

Once the warp map is obtained, the distorted geometry can be produced 
by several methods, including:
\begin{itemize}
\item Creating a cubic environment using a perspective projective onto
4 sides of a cube, which is then resampled onto an angular fish-eye 
image prior to warping. This approach is best for movie-style content,
where high image quality is required; or
\item Mapping directly to the warp map from the cubic environment by 
modifying the texture coordinates, without the need for the intermediate 
fish-eye step.  This is ideal for interactive or real-time data exploration, 
and we have implemented this approach with OpenGL applications. 
\end{itemize}

There is a variation in the light path to different parts of the dome, 
that causes an uneven brightness across the dome. This is corrected 
by applying a non-uniform gradient across the image.  The one form 
of distortion that cannot be corrected for is the need for variable focus 
across the mirror.  This is not a major problem if a projector and lens 
combination is chosen with a good depth of focus, and the front/centre of 
the image is projected with the region of sharpest focus.

With our mirror solution, the full dome surface is not illuminated. This 
is intentional and is similar to most fish-eye projection solutions which 
project onto 3/4 of the dome surface. A dual projector arrangement 
with a single edge blend across the centre is the simplest way to get 
complete dome coverage. 

With a very basic set-up consisting of a laptop (running Mac OS-X), an XGA
projector and $1/4$ spherical mirror, we have 
successfully tested our mirror system in a range of dome sizes: from 
a 3 m rigid, upright dome, 5 m diameter inflatable domes 
to 11 m diameter fixed domes.  In a side-by-side
comparison with a commercial fish-eye solution, 
there was no significant difference
in the projected images.  The image quality depends on the type of content 
that is being viewed, and as with all single-projector solutions using 
XGA projectors, it is hard to obtain good point sources.  Our early 
testing showed that there was substantial ghosting from using a 
back-silvered plexiglass mirror.  This effect was removed by using a 
front-surface mirror---a chrome coating was applied to the plexiglass. 
However, this surface is much more delicate and must be treated with care.  

\section{Three-Dimensional Displays}
\label{3D}
Although a useful intermediary tool, displaying three-dimensional data on
a two-dimensional monitor cannot always provide a full understanding 
of a dataset.

In a spectral line cube, structures may extend beyond one slice, yet a 2D 
display often requires the user to remember what other slices looked in order
to build a mental picture of the 3D distribution.  An improvement is to use
volume rendering [for descriptions of the technique and astronomical 
applications, see Drebin, Carpenter, \& Hanrahan 1988; Gooch (1995); 
Oosterloo (1995); Beeson et al.\ (2003), and Rixon et al.\ (2004)]
or isosurfaces, creating a 3D object out of the
data that can be interactively rotated and examined.  Combined with
lighting, textures and shading this produces a very realistic image, 
but with an assumption that we can understand the type of abstract 3D
structure that is viewed (Rots 1986).   

\subsection{Stereoscopic Projection}
A stereoscopic image is produced by presenting different views to the
left and right eye---changes in the horizontal parallax of foreground
and background objects result in a perception of depth.  Various techniques
exist for presenting stereoscopic images, however, we restrict our discussion 
below to techniques that can be used for large-scale projection of digital
content, suitable for collaborative visualisation or public presentation, with 
real-time interaction.  

Perhaps the most well-known solution are red/blue or red/green 
anaglyph glasses, which are cheap and easy to produce. 
Chroma-Depth$^{\small{TM}}$ glasses
were developed by Steenblik (1996) and use a pair of prisms to disperse and
then recombine light, such that colour provides parallax information:  
red objects appear closer to the viewer than blue objects.  
Thin lines and low line or point densities are required for the best effects, 
and this approach is less effective for isosurfaces  [e.g. Verwichte 
\& Galsgaard (1998), who used chromo-stereoscopy to present simulations 
of prominence formation, however, for an application using 
isosurfaces effectively to study the large-scale structure of the
Universe, see Hultquist et al.\ (2003)].  As with anaglyphs, 
chroma-stereoscopic images can be presented on monitors or printed as 
hardcopies.  The main limitation of these two approaches is the lack 
of colour for anything other than depth information.

A full-colour approach is to use a single projector operating 
at a higher than normal refresh rate (e.g. 120 Hz) that alternately 
displays left and right images.  The images are viewed using electronic 
glasses that switch between transparent and opaque for each eye, 
synchronised to the projector.   While it may appear that this method
would have no cross-talk between the two images, this is not case.
The combined effects of switching time, phosphor decay (for CRT projectors), 
and the scan-line pattern mean that significant ghosting can occur. We have
found that this approach results in the most eye-strain over 
extended periods of usage, most likely due to the flickering of the 
shutter glasses.  In addition, the glasses themselves
are heavy to wear (compared to plastic or even cardboard-framed glasses
that other methods utilise) and can be quite expensive and fragile. 

In our experience, crossed polarising filters provide one of the most 
effective passive stereoscopic methods.  Two data projectors, producing one 
image for each eye, are equipped with linear or circular filters.  The 
viewer wears polarising glasses with
filters that match those of the projector. The advantage is that full-colour
images can be displayed, providing a much more vivid and realistic stereo
environment.   
An additional hardware requirement is a polarisation preserving 
screen, as a normal screen such as a painted 
wall\footnote{We have performed some initial experiments with 
various metallic paints to create a low-cost polarisation-preserving 
surface.  Although the image gain is lower, and cross-talk is higher 
than for commercial screens, a three-dimensional image is visible.} 
will depolarise the incident
light.  Both left and right eye images are projected simultaneously, 
using two outputs from the graphics pipe of a single machine.  The main 
disadvantage for linear filters is that the audience members cannot tilt 
their heads by more than a few degrees.  Apart from this, the amount of 
cross-talk or ghosting for linear filters is minimal, becoming more 
noticeable for high contrast images. This situation can be partly improved by 
using circular filters, but typically at higher cost for both filters 
and glasses, and we have found that there is more overall ghosting. 

\begin{figure*}[htb]
\includegraphics[width=15cm]{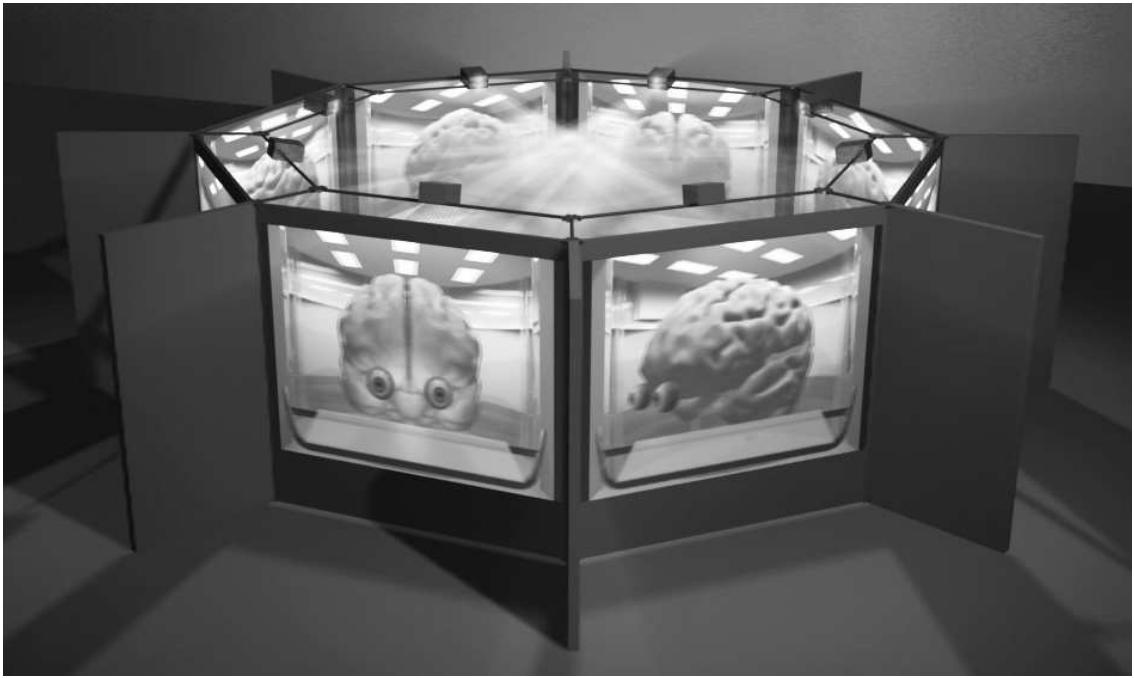}
\caption{The Virtual Room, an 8-wall, rear-projected stereoscopic system
that the audience can walk around in order to see a `contained' version
of a dataset.  Image by E.Hallein.} 
\label{vroom}
\end{figure*}

There are two additional variations on the techniques outlined above: 
front versus
rear projection.  In the former case, the projector is on the same side of 
the screen as the audience. This often requires that the projectors are mounted
up high, which can lead to additional expense and lack of portability.  
For rear projection, the projector(s) are behind the screen, on the 
opposite side to the audience, which means that the projection system requires
additional physical space.  For polarising solutions, a rear projection 
surface that maintains polarisation with minimal loss due to absorption is
required---in general, the screen material is very different for front
and rear projection.

The simplest stereoscopic projection environment is a single, flat wall.  
A multiple wall environment comprises two or more screens, with a range
of angles between the walls.  In all multiple wall environments, it is often
necessary to nominate a `sweet spot' where the viewer should be located.  When
creating stereoscopic content, either computer-generated or through 
photography, etc. knowledge of the viewer position is required.  For a single
wall system, moving away from the preferred position (either towards or
away from the screen, or off-axis parallel to the screen) results in a
distortion of the stereoscopic projection.  The situation is much more 
complex for multiple walls, so these environments are often best-suited 
to a single viewer. 

A side-by-side multiple-wall environment is an extension of the 
tiled projector situation discussed in Section \ref{tile}, 
with a requirement for
edge-blending (now making sure there is consistency between both left and
right pairs of images).    Angling the walls provides a more immersive
environment, and when combined with head-tracking, enables
the viewer a greater range in the directions they can look and move.  
A further extension are the CAVE-style environments (Cruz-Neira et al.\ 1993), 
where there are usually five rear-projected walls: front, two sides, 
roof, and floor.
In some cases, a sixth screen is added at the back, totally enclosing the 
viewer.
Another option for multiple-walls are curved screens, such as in the 
SGI Reality Center,\footnote{http://www.sgi.com/products/visualization/realitycenter} which typically uses 3 edge-blended projectors, or the Advanced
Visualization and Interaction Environment (AVIE) developed at iCinema
(University of New South Wales).  AVIE is an immersive environment, 10 m
in diameter and 4 m high, that surrounds the audience with a 360$^\circ$ 
stereoscopic panorama. 

\subsection{The Virtual Room}
The Virtual Room is an 8-wall, rear-projected stereoscopic system, as 
shown in Figure \ref{vroom}.\footnote{The construction 
of the Virtual Room at the Melbourne Museum 
was funded by a Victorian State Government through the Science \& Technology 
Initiative (STI) grant scheme.  It represents a collaboration between 
Swinburne University of Technology, RMIT, Monash University, Museum Victoria, 
Adacel Pty Ltd. and the University of Melbourne. See http://www.vroom.org.au}  
Unlike other multiple-wall stereo systems, where the viewer is placed inside a 
virtual space, 
the Virtual Room can be thought of as a virtual container---the viewer 
stands on the outside and is able
to walk around the Virtual Room in order to obtain different perspectives.

The Virtual Room is a collaborative environment, suitable for $\sim60$ people
to experience at one time.   As the most expensive advanced display 
that we have constructed, we would not propose that every astronomer 
research group needs to purchase one.  However, we can envision its use during
a workshop, perhaps for a collaborative investigation of multi-wavelength
data.  Each stereoscopic screen could show a specific wavelength or 
simulation, and researchers could move from screen to screen exploring and
discussing their results. 

\subsection{Head-Mounted Displays}
Head-mounted displays (HMDs) provide a near-complete immersive 
experience---the wearer receives a view of their data without 
distractions from the
environment, although the HMD itself can be quiet distracting.   
The major drawbacks of this approach are: 
\begin{itemize}
\item Low resolution: while some of the more expensive models have resolutions
of 1280 x 1024 per eye, most devices are much lower resolution: 640 x 480 and 
800 x 600 being quite common;
\item Awkward to wear: newer devices are much lighter, but they are still
intrusive to wear.  Although we have not used HMDs, our experience of 
light-weight plastic polarising  glasses versus heavier electronic shutter 
glasses has convinced us that lighter is better when choosing hardware that
is worn for extended periods; 
\item Eye fatigue: studies have indicated some critical side-effects of
using HMDs, including nausea, severe vision problems and motion sickness 
[e.g. Geelhoed, Falahee, \& Latham (2000) and references therein].
\end{itemize}

\subsection{Autostereoscopic Displays}

An autostereoscopic display allows the user to see a stereoscopic view
without the need for glasses. One approach is to use a lenticular screen
placed in front of (or integrated into) an LCD monitor.  Reducing the
overall resolution of the monitor, the lenticular gratings direct alternating
vertical lines to the left and right eye.  Alternatives include
using layers of LCD panels, or swept volume displays where images are 
projected onto a rotating blade and the persistence of vision of the 
viewer's eye causes a `solid' object to appear that can be viewed from 
a wide range of angles.

While advanced image displays continue to require specific processing
environments, specialised rooms, or intrusive stereo glasses, they
run the risk of being underused.  
The autostereoscopic display shows a great deal of promise in the years ahead,
as models that also operate like a conventional 2D monitor can easily be
integrated onto the astronomer's desk.   Taking the advanced display to the
astronomer is preferable to taking the astronomer to the advanced display.

\section{The AIDA Survey}
\label{aida_survey}
From 7 March to 2 May 2005, we conducted the Advanced Image Display
in Astronomy (AIDA) survey.   Advertised to members of the Astronomical
Society of Australia (ASA), this web-based survey was designed to provide 
a snapshot of the level of awareness of advanced image displays amongst 
the society's membership.\footnote{As of May 2005, there were 432 ASA members.
See \mbox{http://www.atnf.csiro.au/asa\_www/asa.html}. } The AIDA survey 
received 41 responses, or just under 10\% of the ASA membership.   Due to 
the low response rate,  the results should not be taken as indicative of the 
wider astronomy  research community either within Australia or internationally. 
However, the survey has raised several issues that are worthy of comment, 
and that are guiding our on-going work in this field.

The AIDA survey comprised fourteen questions requiring simple box-ticking 
responses, the fifteenth question was an opportunity to provide general 
comments on advanced displays.  Further details of the questions and 
responses may be found in the Appendix.

The first set of questions (Q1-3) were used to look at the demographics of 
the sample.  We received responses from 17 Masters/PhD students, 10 
postdoctoral fellows, 8 tenured/permanent academics and 4 researchers 
in contract positions.  There was also one response from an undergraduate 
student and one retired academic.

The next set of questions (Q4-9) looked at the astronomical interests and 
current visualisation approaches.  12 respondents identified themselves 
as radio astronomers, 12 as optical astronomers and 8 as computational 
astronomers (most of these were students, perhaps indicating the growth in 
this field within Australia in recent years).

Presented with a list of standard visualisation tools (including additional 
tools that were suggested by the respondents) we found that the astronomers 
in our sample were using an average of 3.2 visualisation
and analysis tools each.  The trend in using packages such as \textsc{Iraf} 
and \textsc{Miriad} was consistent with the number of optical and radio 
astronomers in the sample.  For advanced image displays to be 
useful and usable, they must be compatible with a wide range of 
packages and data-formats.

Custom
\textsc{PGPlot}\footnote{\mbox{http://www.astro.caltech.edu/$\sim$tjp/pgplot/}} tools were also widely used (44\% of respondents).  This demonstrates the willingness of astronomers to write their own code when existing tools are not capable of producing all of their analysis and visualisation needs.  Awareness
of advanced displays needs to be supported with awareness of programming techniques, such as a set of basic
\textsc{PGPlot} routines that are compatible with stereoscopic projection, digital domes or other display types.  We are now developing such a set of programming tools.

All of our respondents indicated that they visualise their data in some form, whether it be simple graphs, histograms and plots, or 2D images.   While 11 respondents had used 3D images, only 6 had actually used a three-dimensional display technique, such as stereoscopic projection.

We were somewhat surprised by the response to question 7: How would you 
describe the dimensionality of the majority of your data?   Given the 
choice of either one, two, three or N-dimensional, only half
of the respondents indicated that they used data with $N \geq 3$ dimensions.  
It would seem that the respondents were only considering Cartesian 
coordinates, rather than also counting other parameters
as contributing to the dimensionality of the dataset, and this outcome 
may have been affected by the way the question was posed.  
Consider a computational example: a typical N-body simulation contains
three dimensional particle positions, but also calculates parameters such as 
velocity, density, gravitational potential, etc. in a time-varying fashion.  
Such a dataset has a dimensionality $N > 3$, and any subset of these 
parameters could be visualised and explored in order to gain a
better understanding of their relationships, particularly with one of the 
three-dimensional displays we discussed in section \ref{3D}.  
If astronomers are not thinking of their datasets in this way, we would 
encourage them to start.

The third set of questions (Q10-14) was aimed at testing the level of awareness of specific advanced image displays, including whether respondents had used these devices, whether they saw a benefit from them, and what aspects might be preventing their uptake.  Only 6 respondents had ever used an advanced image display in their astronomy research.   While only 16 people indicated that they saw a definite benefit from using advanced displays,  none responded `definitely no' (the remainder
of the sample selected `perhaps'). We should not draw too many conclusions from this: we cannot test whether those astronomers that took part in the AIDA survey are the only ones from the ASA membership that see some benefit from using advanced image displays.  We remain hopeful that with growing awareness of the available tools and techniques, more astronomers will see some benefit to their work.

The most informative result to come out of the AIDA survey were the reasons why our respondents were not currently making use of advanced image displays.  The most common limiting factor (selected by $73\%$ of the sample) was lack of knowledge of advanced displays, followed by lack of software tools and access to local facilities ($46\%$ each).  Along with the choices we proposed, individuals indicated other factors such as lack of time to investigate advanced displays or develop software for them, lack of knowledge of available software, and medical/physiological conditions.\footnote{Some advanced displays are not appropriate for all users.  For example, stereoscopic 3D displays are not very useful for the $\sim10\%$ of the population who do not have binocular vision.  Headaches can arise from overuse of particular techniques, especially when the eyes are forced to focus in an unnatural manner for extended periods of time. For further comments on this issue, see Rots (1986).}

The final question was an opportunity for the participants to share any other thoughts they had about the usefulness or otherwise of advanced image displays, types of advanced displays they would be interested in learning more about, etc.  We present a selection of the responses below:
\begin{itemize}
\item {\em I doubt} [the] {\em usefulness if it is not readily available every 
time I need to visualise data.}
\item {\em Needs to be accessible to me personally and easily programmed/operated.  }
\end{itemize}We agree that easy access to advanced image displays is important if they
are to be widely used.  This can be achieved most successfully through local
facilities, but these require local expertise to go with them.  The need for
these systems to be easily programmed perhaps goes back to the wide use of
custom code (e.g. \textsc{PGPlot}) written by astronomers, and may be seen more as a requirement for the development of a set of standard visualisation tools.
However, any generic tool is unlikely to meet the needs of every astronomer.

\begin{itemize}
\item {\em These displays definitely help in the visualisation of many
different phenomenon and I have seen [them] used successfully in a number
of different instances}...[however]...{\em I'd be more interested in
knowing more about the tools or methods required to make use of these displays.}
\item {\em ...training would be essential to get people over the threshold to start
using these facilities}
\item {\em For my own research the 2D display showing the 3D data has been sufficient there has never really been the need to walk down the hall to the Virtual Reality room.}
\end{itemize}
Awareness of advanced displays and the tools that go with them seems to be
one of the big contributors to the current lack of uptake.  As facilitators
of advanced displays, we need to work harder to help astronomers through the
first few steps in using these devices so that researchers can judge for
themselves whether there are benefits from looking at their data in new ways.
\begin{itemize}
\item {\em I don't really see any benefit to using these tools in my
research---but
they would be very valuable in teaching.  I have not, however, used them for this due to the amount of my time needed to figure out how to do it.}
\end{itemize}
Swinburne University of Technology has had great success in using stereoscopic
displays for public education activities, using the approach that the same
tools we use to view astronomical datasets for research are equally applicable
for educational purposes.  We would suggest that any institution or
research group that installs an advanced display for research automatically
has a valuable teaching tool at their disposal.

\section{The Limits of Visualisation}
Norris (1994) identified four important features of a visualisation system.
They should:
\begin{enumerate}
\item Allow the user to gain an intuitive understanding of the dataset;
\item Let the user see features in the data that would not be obvious
using other approaches;
\item Help the user to get quantitative results; and
\item Enable results (both qualitative and quantitative) to be communicated
to others.
\end{enumerate}

The advanced display is only one part of a complete visualisation system,
and the usefulness of the displays will always be limited by the availability
of software tools and the capability of the hardware that drives the
display. For large datasets, the response speed/latency can limit the
effectiveness of interactivity (e.g. Welling \& Derthick 2001). 
There is a need for software tools that work consistently on range of 
displays (Rots 1986; Mann et al. 2002), so that users do not have 
constantly switch between data formats and user interfaces (Brugel et al.
1993).  An example of where this software scalability has been successful 
is Partiview,\footnote{http://niri.ncsa.uiuc.edu/partiview/} a cross-platform 
application that works on laptops, desktops and the 21 m Hayden 
Planetarium dome (Levy 2003).

One of the major challenges facing advanced image displays is the
lack of hardcopies---if a researcher cannot print out the results of
an investigation and publish it in a journal or send it to another
researcher for comment, is it worth the effort?  We note that similar 
problems still exist with regards to publishing videos or animations, and even 
high-resolution images in colour in some scientific journals.
The growing move towards electronic publication means that 
datasets can be shared more easily over the Internet.  Combine this with
a standard data format and greater availability of local facilities, 
and the lack of hardcopies might not be such a limitation in the future.  

To date, our own software effort has focused 
on qualitative data exploration, as these can be implemented without
sophisticated user interaction devices (e.g. a mouse or keyboard commands
can be used to perform simple tasks like rotating datasets, zooming in and
out, etc.).   Quantitative tools can require a higher level of sophistication
as part of the user interaction---is it intuitive to identify and
highlight a three-dimensional region with a series of key-presses?  While
interaction devices exist (e.g. 3D `mice' that can measure multiple degrees of 
freedom and the electronic `wand' used in the CAVE), they often take some
practice to use effectively.  A more natural approach might be to let the
user simply point with their hands at a particular region or object. 
Multi-sensory data exploration, where the astronomer is provided
information via the senses of sight, hearing and touch, offers some 
tantalising ideas for the way future astronomers might interact with, and
immerse themselves within, their data.

\section{Final Thoughts}
Along with our goal of obtaining a snapshot of the level of 
awareness and uptake of advanced imaged displays amongst members 
of the Astronomical Society of Australia, the AIDA survey was intended to help astronomers 
start to think about the ways that advanced displays could help them with their research work. 

Traditional two-dimensional displays (paper, monitors, overhead projectors) 
will always remain incredibly valuable, and we are not attempting to 
suggest that they should be replaced.  Many visualisation tasks 
can be accomplished with contour plots, graphs, histograms. What we wish 
to emphasise is that with the aid of advanced display technology, visualisation 
can go well beyond this.  To paraphrase Rots (1986): `One should not 
imagine that display tools come for free', however, advanced image 
displays are an affordable reality, and today's advanced display 
may well be tomorrow's commonplace system.  

We leave the close-to-final word on the AIDA survey to one of 
our participants, as it summarises many of our own views on this subject:
{\em `We have now reached a point, with large-scale surveys 
and multi-wavelength databases, that we really need to be able to 
visualise multi-dimensional datasets to advance our understanding, but in 
general institutions often lack either appropriate software or display 
equipment and so individuals revert to traditional methods of display. 
There is also a time pressure that acts to prevent overworked researchers 
from learning new technologies as they simply [cannot] commit the hours 
required to learn them which hinders the desire to investigate possible 
advanced display options even when available.' }

With our on-going work, and supporters out there 
amongst the astronomical community, we hope to see advanced displays in
more regular use in the years that follow.  In the short term, the authors 
look forward to helping researchers explore their data with alternative 
approaches, as we have knowledge and experience we would like to share.
It is time for astronomers to think outside the square frames of their monitors,
and truly immerse themselves in the data.

\section*{Acknowledgments} 

This research has been funded by a Swinburne Researcher Development Grant.
We would like to thank David Barnes and Evan Hallein for their valuable
contributions; Glen Moore from the Wollongong Science Centre and 
Charles Treleaven from Cosmodome Australasia for access to their domes 
(fixed and portable);  and Roberto Calati for his assistance with implementing 
edge-blending.  We are appreciative to all of the respondents to 
the AIDA survey.  We thank the anonymous referee for very useful comments
that modified the direction of this paper.


\appendix

\section{The AIDA Questions}
We present in this appendix the questions and answers from the AIDA
survey.  Note that when percentages are summed, they may differ from 
$100\%$ due to rounding.  Each question had an option to provide 
`no answer', and these responses have not been removed from the sample.

\textbf{Q1. What is your current position?} Our sample
consisted of 17 Masters/PhD students, four postdoctoral fellows on their 
first placement, six postdoctoral fellows on their second or later 
placement, eight tenured/permanent academics, four researchers in 
contract positions, one retired academic and one undergraduate student.
To simplifying reporting, we introduce two broad categories of 
ASA members: \textbf{students} (44\% of the sample) and a group that 
might loosely be defined as \textbf{senior} researchers (56\%).  

\textbf{Q2. Where did you complete your most recent degree?}
23 respondents had completed their most recent degree in Australia, 7 
in the UK, 6 in other European countries, 2 in the US, and 1 each in Canada 
and New Zealand.

\textbf{Q3. How recently did you complete your PhD studies?}
With 18 student respondents yet to complete, 8 respondents
had completed their PhD within the last 5 years, 7 had completed
more than 5 but fewer than 10 years ago, and 7 had completed more than
10 years ago.

\begin{table*}[pt]
\caption{AIDA Q4. What are your main astronomical interests (up to three
choices per respondent)?
The number of responses and percentage for the category is
given for each of students (18 responses), senior researchers (23 responses),
and the total sample (41 responses).}
\begin{center}
\begin{tabular*}{\textwidth}{@{\extracolsep{\fill}}lrlrlrl}
\label{t_q4}
\textbf{Research area}
& \multicolumn{2}{c}{\textbf{Students}}
& \multicolumn{2}{c}{\textbf{Seniors}}
& \multicolumn{2}{c}{\textbf{Total}} \\

Galaxies (e.g. formation, evolution) & 12 & (67\%) & 13 & (57\%) &  25 & (61\%) \\
Cosmology & 5 & (28\%) & 11 & (48\%) & 16 & (39\%) \\
Stars (e.g. formation, evolution, structure)  & 6 & (33\%) & 6 & (26\%) & 12 & (29\%) \\
Stellar clusters & 4 & (22\%) & 4 & (17\%)  & 8 & (20\%) \\
Milky Way and/or Local Group & 4 & (22\%)& 4 & (17\%) & 8 & (20\%) \\
Quasars & 1 & (6\%)& 6 & (26\%) & 7 & (17\%) \\
Supernovae, Pulsars, Black Holes or other stellar remnants & 2 & (11\%)& 3 & (13\%) & 5 & (12\%)\\
Instrumentation & 3 & (17\%)& 2 & (9\%) & 5 & (12\%) \\
Planets (e.g. formation, evolution, structure) & 1 & (6\%) & 2 & (9\%) & 3 & (7\%)\\
Nebulae & 1 & (6\%) & 0 & (0\%)) & 1 & (2\%) \\
Interstellar Medium & 0 & (0\%) & 1 & (4\%) & 1 & (2\%)\\
Virtual Observatory & 1 & (6\%) & 0 & (0\%) & 1 & (2\%)\\
\multicolumn{1}{r}{\textbf{Total}} & 40 & &  52 & & 92 \\

\end{tabular*}
\end{center}
\end{table*}

\begin{table*}[pt]
\begin{center}
\caption{AIDA Q5. What is your main role?
The number of responses and percentage for the category is
given for each of students (18 responses), senior researchers (23 responses),
and the total sample (41 responses).}
\medskip
\begin{tabular*}{0.7\textwidth}{@{\extracolsep{\fill}}lrlrlrl}
\label{t_q5}
\textbf{Main role}
& \multicolumn{2}{c}{\textbf{Students}}
& \multicolumn{2}{c}{\textbf{Seniors}}
& \multicolumn{2}{c}{\textbf{Total}} \\
Radio astronomer & 6 & (33\%) & 6 & (26\%) & 12 & (29\%) \\
Optical astronomer &  4 & (22\%) & 8 & (35\%) & 12 & (29\%) \\
Computational astronomer & 6 & (33\%) & 2 & (9\%) & 8 & (20\%) \\
Infrared astronomer & 0 & (0\%) & 1 & (4\%) & 1 & (2\%) \\
Multiwavelength astronomer & 1 & (6\%) & 2 & (9\%) & 3 & (7\%) \\
Theorist & 1 & (6\%) & 4 & (17\%) & 5 & (12\%)
\end{tabular*}
\end{center}
\end{table*}

\begin{table*}[pt]
\caption{AIDA Q6. How would you describe the majority of the data that you use?
The number of responses and percentage for the category is
given for each of students (18 responses), senior researchers (23 responses),
and the total sample (41 responses).}

\begin{center}
\begin{tabular*}{0.95\textwidth}{@{\extracolsep{\fill}}lrlrlrl}
\label{t_q6}
\textbf{Data size and type}
& \multicolumn{2}{c}{\textbf{Students}}
& \multicolumn{2}{c}{\textbf{Seniors}}
& \multicolumn{2}{c}{\textbf{Total}} \\Large-scale numerical simulation ($> 10^8$ particles) & 0 & (0\%) & 1 & (4\%) & 1 & (2\%) \\
Medium-scale numerical simulation ($10^4 < n < 10^8$ particles) & 1 & (6\%) & 3 & (13\%) & 4 & (10\%) \\
Small-scale numerical simulation ($< 10^4$ particles) & 1 & (6\%) & 0 & (0\%) & 1 & (2\%) \\
Large survey ($<50\%$ of sky, $>1000$ objects)&  6 & (33\%) & 6 & (26\%) & 12 & (29\%) \\
Medium survey ($100 < n < 1000$ objects)&  0 & (0\%) & 7 & (30\%) & 7 & (17\%) \\
Small survey ($2 < n < 100$ objects) & 7 & (39\%) & 5 & (22\%) & 12 & (29\%) \\
Single object & 1 & (6\%) & 1 & (4\%) & 2 & (5\%) \\
\end{tabular*}
\end{center}
\end{table*}

\begin{table*}[htb]
\begin{center}
\caption{AIDA Q8. Which tools do you regularly use to analyse your data?
The number of responses and percentage for the category is
given for each of students (18 responses), senior researchers (23 responses),
and the total sample (41 responses).}
\medskip
\begin{tabular*}{0.7\textwidth}{@{\extracolsep{\fill}}lrlrlrl}
\label{t_q8}
\textbf{Analysis tool}
& \multicolumn{2}{c}{\textbf{Students}}
& \multicolumn{2}{c}{\textbf{Seniors}}
& \multicolumn{2}{c}{\textbf{Total}} \\
\textsc{Miriad} & 8 & (44\%) & 10 & (43\%)& 18 & (44\%) \\
Custom \textsc{PGPlot} tools & 6 & (33\%) & 12 & (52\%) & 18 & (44\%) \\
\textsc{Iraf} & 6 & (33\%) & 11 & (48\%) & 17 & (41\%) \\
\textsc{Karma} & 7 & (39\%)& 9 & (39\%)& 16 & (39\%) \\
\textsc{Mongo}/\textsc{Supermongo} & 5 & (28\%)& 9 & (39\%) & 14 & (34\%) \\
\textsc{Idl} & 5 & (28\%) & 7 & (30\%) & 12 & (29\%) \\
Other locally developed tool & 5 & (28\%)& 6 & (26\%) & 11 & (27\%) \\
\textsc{Aips} & 3 & (17\%) & 5 & (22\%) & 8 & (20\%) \\
\textsc{Aips++} & 3 & (17\%) & 3 & (13\%) & 6 & (15\%)  \\
Other commercially developed tool & 2 & (11\%) & 4 & (17\%) & 6 & (15\%)\\
\textsc{Matlab} & 3 & (17\%) & 1 & (4\%) & 4 & (10\%) \\
Microsoft \textsc{Excel} & 1 & (6\%) & 1 & (4\%) & 2 & (5\%) \\
\multicolumn{1}{r}{\textbf{Total}} & 54  & & 78 &  & 132
\end{tabular*}
\end{center}
\end{table*}

\begin{table*}[htb]
\begin{center}
\caption{AIDA Q9. What are the main ways you visualise your data?
The number of responses and percentage for the category is
given for each of students (18 responses), senior researchers (23 responses),
and the total sample (41 responses).}
\medskip
\begin{tabular*}{0.75\textwidth}{@{\extracolsep{\fill}}lrlrlrl}
\label{t_q9}
\textbf{Visualisation method}
& \multicolumn{2}{c}{\textbf{Students}}
& \multicolumn{2}{c}{\textbf{Seniors}}
& \multicolumn{2}{c}{\textbf{Total}} \\

Graphs, histograms, plots & 18 & (100\%) & 21 & (91\%) & 39 & (95\%) \\
2D images & 16 & (89\%) & 19 & (83\%) & 35 & (85\%) \\
3D images & 5 & (28\%) & 6 & (26\%)& 11 & (27\%) \\
OpenGL interaction & 1 & (6\%) & 0 & (0\%) & 1 & (2\%) \\
Animations & 0 & (0\%) & 1 & (4\%)  & 1 & (2\%)
\end{tabular*}
\end{center}
\end{table*}

\begin{table*}[htb]
\begin{center}
\caption{AIDA Q12. What is your experience of selected advanced image displays?
In this table, awareness is rated on a scale from A--E$^a$.  Results are
given for the categories of students (18 responses) and seniors
(23 responses), noting that one student selected the `no answer'
option for all display types.  A distinction can be made
between advanced displays that respondents have used (A = regularly,
B = occasionally), are aware of but have not used (C = seen but not used,
D = know what it is but have not seen it in operation), and are unfamiliar
with (E). }
\label{t_q12a}
\begin{tabular*}{\textwidth}{lrr|rr|r||rr|rr|r}
 & \multicolumn{5}{c}{Students} &
 \multicolumn{5}{c}{Seniors} \\
\textbf{Advanced Display} & \textbf{A} & \textbf{B} &
\textbf{C} & \textbf{D} & \textbf{E} &
\textbf{A} & \textbf{B} &
\textbf{C} & \textbf{D} & \textbf{E}  \\
CRT/LCD Monitor &  16 & 0 & 1 & 0 & 0 & 21 & 0 & 2 & 0 & 0\\
Digital dome projection (e.g. full-dome planetarium)  & 0 & 0 & 12 & 5 & 0 & 0 & 2 & 18 & 2 & 1 \\
Multiple projector tiled display  & 0 & 2 & 8 & 4 & 3& 0 & 2 & 10 & 3 & 8\\
Stereoscopic projection (single screen)  & 0 & 3 & 10 & 3 & 1&1 & 2 & 14 & 4 & 2\\
Multiple wall stereo projection (2 or more walls)  & 0 & 0 & 5 & 8 & 4& 0 & 0 & 6 & 11 & 6\\
Curved stereoscopic environment (e.g. Reality Centre)  & 0 & 0 & 2 & 9 & 6 & 0 & 0 & 6 & 10 & 7\\
Head-mounted display  & 0 & 0 & 4 & 8 & 5 &0 & 0 & 6 & 11 & 6\\
Autostereoscopic display  &0 & 0 & 1 & 3 & 13 & 0 & 0 & 1 & 7 & 15 \\
The Virtual Room  & 0 & 0 & 1 & 5 & 11 & 0 & 0 & 3 & 12 & 8
\end{tabular*}

\medskip
$^a$ See Q12 for a full description of the symbols A--E.
\end{center}
\end{table*}

\begin{table*}[htb]
\begin{center}
\caption{AIDA Q13. Which factors (if any) are currently preventing you
from using advanced image displays in your research?
The number of responses and percentage for the category is
given for each of students (18 responses), senior researchers (23 responses),
and the total sample (41 responses).}
\medskip
\begin{tabular*}{0.75\textwidth}{@{\extracolsep{\fill}}lrlrlrl}
\label{t_q13}
\textbf{Limitations}
& \multicolumn{2}{c}{\textbf{Students}}
& \multicolumn{2}{c}{\textbf{Seniors}}
& \multicolumn{2}{c}{\textbf{Total}} \\
Lack of knowledge of available displays  &  13 & (72\%) & 17 & (74\%) & 30 & (73\%) \\
Lack of software tools & 8 & (44\%) & 9 & (39\%) & 19 & (46\%) \\
Lack of local facilities  &  10 & (56\%) & 9 & (39\%) & 19 & (46\%) \\
Cost of advanced image displays & 7 & (39\%)  & 10 & (43\%) & 17 & (41\%) \\
Lack of national facility & 4 & (22\%) & 5 & (22\%) & 9 & (22\%) \\
Other & 4 & (22\%) & 6 & (26\%) & 10 & (24\%)
\end{tabular*}
\end{center}
\end{table*}

\begin{table*}[t!]
\begin{center}
\caption{AIDA Q14. If you have seen a stereoscopic projection system in action, where was it?
The number of responses and percentage for the category is
given for each of students (18 responses), senior researchers (23 responses),
and the total sample (41 responses).}
\medskip
\begin{tabular*}{0.75\textwidth}{@{\extracolsep{\fill}}lrlrlrl}
\label{t_q14}
\textbf{Location}
& \multicolumn{2}{c}{\textbf{Students}}
& \multicolumn{2}{c}{\textbf{Seniors}}
& \multicolumn{2}{c}{\textbf{Total}} \\

Swinburne University &  9 & (50\%)  & 9 & (39\%) & 18 & (44\%) \\
Parkes Observatory & 2 & (11\%) & 3 & (13\%) & 5 & (12\%) \\
Sydney Observatory & 2 & (11\%) & 1 & (4\%) & 3 & (7\%) \\
Jodrell Bank Observatory & 0 & (0\%) & 1 & (4\%) & 1 & (2\%) \\
Australian National University & 1 & (6\%) & 2 & (9\%) & 3 & (7\%) \\
Another institution in Australia &  1 & (6\%) & 3 & (13\%) & 4 & (10\%) \\
Another institution overseas & 1 & (6\%) & 4 & (17\%) & 5 & (12\%) \\
Tradeshow (in Australia) & 0 & (0\%) & 1 & (4\%) & 1 & (2\%) \\
Tradeshow (overseas) & 1 & (6\%) & 1 & (4\%) & 2 & (5\%)
\end{tabular*}
\end{center}
\end{table*}

\textbf{Q4. What are your main astronomical interests?} 
Participants were able
to select up to three research areas from a list, or provide their own choice.
Table \ref{t_q4} shows the results, with responses separated into student and senior
groups.  

\textbf{Q5. What is your main role?}
For this question, respondents were asked to nominate their main role
from a list of options, or to propose an alternative.  A summary of 
the responses for the student and senior groups is given in Table \ref{t_q5}.   

\textbf{Q6. How would you describe the majority of the data that you use?}
This question gave respondents a choice of nine different data sample sizes,
including small to large numerical simulations, small to all-sky surveys,
single object or instrument design, with results in Table \ref{t_q6}.  

\textbf{Q7. How would you describe the dimensionality of the majority of
your data?}
Due to the possible confusion over the wording in this question based
on the results received, we do not present a detailed breakdown.   See
Section \ref{aida_survey} for a discussion.

\textbf{Q8. Which tools do you regularly use to analyse your data?}
Participants were presented with a range of standard data reduction,
analysis and visualisation packages.  Individuals reported using
between zero and eight different packages. A summary of results is shown 
in Table \ref{t_q8}.  

\textbf{Q9. What are the main ways you visualise your data?}
Participants were presented with a range of visualisation methods.
On average, astronomers used $\sim2$ different methods each to visualise 
their data.  Results are in Table \ref{t_q9}. 

\textbf{Q10. Have you ever used an advanced image display for astronomy
research?}
Only 6 respondents (15\%) reported having used an advanced image display
in their astronomy research: 1 student and 4 seniors.

\textbf{Q11. Do you see a benefit from using advanced image displays for
astronomy research?}
Participants were given a choice of three answers.  16 respondents
(39\%) selected `yes' and 25 selected `perhaps'. No respondents
selected `definitely no'.

\textbf{Q12. What is your experience of selected advanced image displays?}
In this question, we identified one common class of 2D image display (CRT 
or LCD
monitors) and eight advanced image displays.  Participants were asked to
rate their knowledge using the following scheme:
\begin{itemize}
\item A: Use the device $> 50\%$ of the time;
\item B: Use the device $< 50\%$ of the time;
\item C: Have seen the device in operation (not just in astronomy),
but have not used it;
\item D: Have not seen the device in operation, although know what it is; and
\item E: Not familiar with the device.
\end{itemize}

Table \ref{t_q12a} gives the number of responses in each category
for the nine different image displays.  We can make a distinction between
A+B and C+D to identify those advanced image displays that the respondents
have used, and those they were aware of without having actually used them.

\textbf{Q13. Which factors (if any) are currently preventing you from using
advanced image displays in your research?}
Five options were presented to participants, with an option to suggest
a limiting factor of their own.  Results are presented in
Table \ref{t_q13}.  

\textbf{Q14. If you have seen a stereoscopic projection system in action,
where was it?}
This question was designed to identify whether there was a `Swinburne Factor'.
As we have been operating a stereoscopic 3D system since 1999, our staff,
students and visitors may have received a much higher level of exposure to
the ideas of advanced image displays than the  broader Australian astronomy
community.  $56\%$ of respondents had not seen the Swinburne 3D theatre in
operation, so it seems that the AIDA survey has reached further than just
our local staff and students.  However, as a fraction of the ASA membership,
we may have received an over-supply of local responses.  Results in 
Table \ref{t_q14}.


\begin{thebibliography}{}

\bibitem[Abbot et al.(2004)]{Abbot04}Abbott, B.~P., Emmart, C.~B., Levy, S., 
Liu, C.~T., 2004, Toward an International Virtual Observatory, ed. P.~J.Quinn 
\& K.~M.Gorski (Berlin: Springer), 57

\bibitem[Beeson, Barnes \& Bourke (2003)]{Beeson03}
Beeson, B., Barnes, D.~G., Bourke, P.~D.,  2003, PASA, 20, 300

\bibitem[Brugel et al. (1993)]{Brugel93} Brugel, E.~W., Domik, G.~O., 
Ayers, T.~R., Final Report Colorado University, Boulder Dept. of 
Computer Science,

\bibitem[Cruz-Neira, Sandin \& DeFanti (1993)]{CruzNeira93} Cruz-Neira, C., 
Sandin, D.~J., DeFanti, T.~A., 1993, Proceedings of the 20th Annual 
Conference on Computer Graphics and Interactive Techniques, 
(New York: ACM Press), 135

\bibitem[Domik & Mickus-Miceli (1992)]{Domik92} Domik, G.~O., 
Mickus-Miceli, K.~D., 1992, Astronomical Data Analysis Software and Systems 
I, ASP Conf. Ser., Vol 25, ed. D.~M.Worrall, C.Biemesderfer, J.Barnes, 95

\bibitem[Drebin et al. (1988)]{Drebin88} Drebin, R.~A., Carpenter, L.~C.,  
Hanrahan, P., 1988, Proceedings of the 15th Annual Conference on Computer
Graphics and Interactive Techniques, ed. R.J.Beach (New York: ACM Press), 65

\bibitem[Fomalont(1982)]{Fomalont82}Fomalont, E.~B., 1982, Synthesis
Mapping, NRAO Workshop Proceedings 5, ed. A.~R.Thompson, L.~R.D'Addario
(Green Bank: NRAO), lecture 11

\bibitem[Geelhoed(2000)]{Geelheod00}Geelhoed, E., Falahee, M., Latham, K., 
2000, LNCS 1927, Handheld and Ubiquitous Computing: Second 
International Symposium, ed. P.Thomas, H.-W.Gellersen (Berlin: Springer-Verlag),
236

\bibitem[Gooch (1995)]{Gooch95} Gooch, R., 1995, ASP Conf. Ser. 77, 
Astronomical Data Analysis Sotware and Systems IV, ed. R.~A.Shaw, H.~E.Payne, 
J.~J.~E.Hayes, 144 

\bibitem[Hultquist et al. (2003)]{Hultquist03} 
Hultquist, C., Perumal, S., Marais, P., Fairall, T., 2003,
Technical Report CS03-16-00

\bibitem[Joye \& Mandel(2004)]{Joye04}Joye, W.~A., Mandel, E., 
2004, Astronomical Data Analysis Software and Systems (ADASS) XIII, ASP
Conf. Proc. 314, ed. F.Ochsenbein, M.~G.Allen, D.Egret (San Francisco: ASP), 505

\bibitem[Levy (2003)]{Levy03} Levy, S., 2003, Astrophysical Supercomputing
Using Particles, IAU Symp. 208, ed. J.Makino, P.Hut (San Francisco: ASP), 343

\bibitem[Mann et al. (2002)]{Mann02}Mann, B., Williams, R., Atkinson, M., 
Brodlie, K., Storkey, A., Williams, C., 2002, Scientific Data Mining, 
Integration and Visualization, Report of the workshop held at the 
e-Science Institute, Edinburgh, 24-25 October 2002 

\bibitem[Murtagh (1989)]{Murtagh89}Murtagh, T., 1989, IrAJ, 19, 17

\bibitem[Norris(1994)]{Norris94} Norris, R.~P., 1994, Astronomical Data Analysis Software and Systems III, ASP Conf. Ser. 61, ed.  D.~R.Crabtree, R.~J.Hanisch, 
J.Barnes, 51

\bibitem[Oosterloo (1995)]{Oosterloo95} Oosterloo, T., 1995, PASA, 12, 215

\bibitem[Rixon et al. (2004)]{Rixon04}Rixon, G., Barnes, D., Beeson, B., 
Yu, J., Ortiz, P., 2004, Astronomical Data Analysis Software and Systems
(ADASS) XIII, ASP Conf. Ser. 314, ed. F.Ochesenbein, M.~G.Allen, D. Egret
(San Francisco: ASP), 509 

\bibitem[Rots(1986)]{Rots86} Rots, A., 1986, Synthesis Imaging, ed. R.A.Perley,
F.R.Schwab, A.H.Bridle (Green Bank: NRAO), 231 

\bibitem[Steenblik (1986)]{Steenblik86}Steenblik, R.~A., 1986, 
US Patent No 4-597-634

\bibitem[Teuben et al.(2001)]{Teuben01} Teuben, P.~J., Hut, P., Levy, S., 
Makino, J., McMillan, S., Portegies Zwart, S., Shara, M., Emmart, C., 2001, 
Astronomical Data Analysis Software and Systems X, ASP Conf. Ser. 238, 
ed.  F.~R.Harnden Jr, F.~A.Primini, H.~E.Payne (San Francisco: ASP), 499

\bibitem[Van Buren et al.(1995)]{VanBuren95}Van Buren, D., Curtis, P., 
Nichols, D.~A., Brundage, M., 1995, Astronomical Data Analysis Software 
and Systems IV, ASP Conf. Ser. 77, ed. R.~A.Shaw, H.~E.Payne, J.~J.~E.Hayes, 99

\bibitem[Verwichte \& Galsgaard (1998)]{Verwichte98} Verwichte, E., 
Galsgaard, K., 1998, SoPh, 183, 445 

\bibitem[Welling \& Derthick (2001)]{Welling01}Welling, J., Derthick, M., 
Virtual Observatories of the Future, ASP Conf. Proc. 225, ed. R.~J.Brunner, 
S.~G.Djorgovski, A.~S.Szalay (San Francisoc: ASP), 284



\end{thebibliography}
\end{document}